\documentclass[sigchi]{acmart}

\usepackage{booktabs} 
\usepackage{balance}

\setcopyright{acmcopyright}
\usepackage{verbatimbox}
\usepackage{tabularx}
\usepackage{subcaption}
\usepackage{animate}
\usepackage{url}

\copyrightyear{} 
\acmYear{} 
\setcopyright{acmcopyright}
\acmConference[Accepted to CHI 2019]{CHI Conference on Human Factors in Computing Systems Proceedings}{May 4--9, 2019}{Glasgow, Scotland Uk}
\acmBooktitle{CHI Conference on Human Factors in Computing Systems Proceedings (CHI 2019), May 4--9, 2019, Glasgow, Scotland Uk}
\acmPrice{15.00}
\acmDOI{}
\acmISBN{}

\usepackage{cuted}
\usepackage{capt-of}

\setcopyright{none}

\begin{document}
\renewcommand\UrlFont{\color{cyan}\small\ttfamily\textbf}
\title[B-Script]{B-Script: Transcript-based B-roll Video Editing with Recommendations}

\author{%
Bernd Huber$\dagger$*, Hijung Valentina Shin*, Bryan Russell*, Oliver Wang*, and Gautham J. Mysore*\\
$\dagger$Harvard University, Cambridge, MA, USA\\
*Adobe Research, USA\\
bhb@seas.harvard.edu, \{vshin,brussell,owang,gmysore\}@adobe.com
}

\newcommand{\val}[1]{\textcolor{blue}{#1}}
\renewcommand{\shortauthors}{Huber B. et al.}

\begin{strip}
    \centering
    \includegraphics[width=0.96\columnwidth]{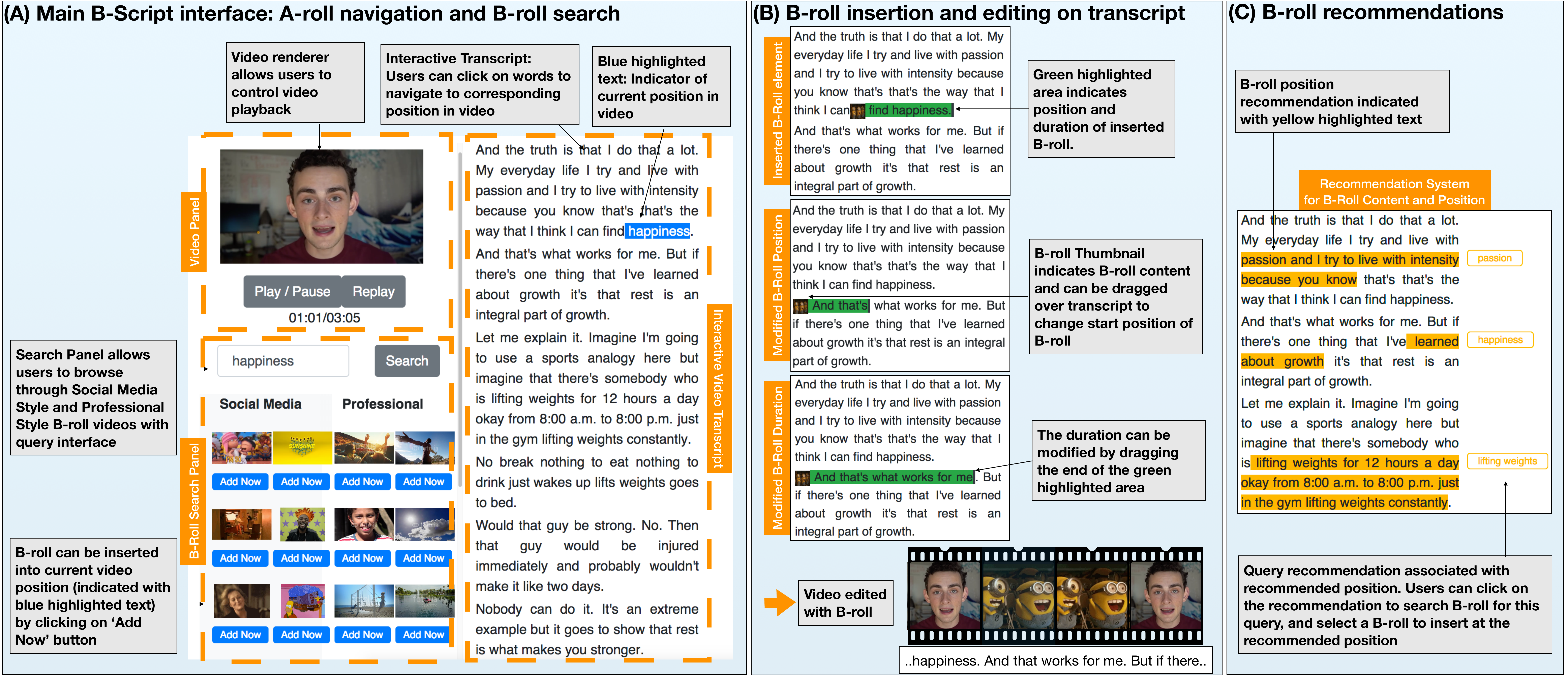}
    \vspace{-8pt}
    \captionof{figure}{Example functionality of our video editing interface, B-Script. (A) Users can navigate the video by clicking on a transcript, and can search for possible B-roll clips from external online repositories. (B) Users can then add B-roll clips with the click of a button, and change their position and duration directly on the transcript. (C) An automatic recommendation system identifies possible locations and search keywords in the video, and presents these to the user for easy edits. A demo of the system can be found at \texttt{\url{berndhuber.github.io/bscript}}.}
    \label{fig:teaser}
    \vspace{-8pt}
\end{strip}

\begin{abstract}
In video production, inserting B-roll is a widely used technique to enrich the story and make a video more engaging. However, determining the right content and positions of B-roll and actually inserting it within the main footage can be challenging, and novice producers often struggle to get both timing and content right. We present B-Script, a system that supports B-roll video editing via interactive transcripts. B-Script has a built-in recommendation system trained on expert-annotated data, recommending users B-roll position and content. To evaluate the system, we conducted a within-subject user study with 110 participants, and compared three interface variations: a timeline-based editor, a transcript-based editor, and a transcript-based editor with recommendations. Users found it easier and were faster to insert B-roll using the transcript-based interface, and they created more engaging videos when recommendations were provided.
\end{abstract}

\maketitle

\section{Introduction}
Video has become arguably the most influential online medium for both personal and professional communication used by millions of people. A common editing technique used to make a video story more engaging is to add visual interest and context by cutting away from the main video (dubbed {\it A-roll}) to secondary video footage (dubbed {\it B-roll}). Note that often the main audio track is not modified and only the visuals are overlaid. B-roll can be taken from secondary or online sources, such as YouTube \footnote{www.youtube.com}, Stock/Getty \footnote{www.gettyimages.com}, or Giphy \footnote{www.giphy.com}. 

\begin{figure}[t]
    \centering
    \includegraphics[width=0.9\columnwidth]{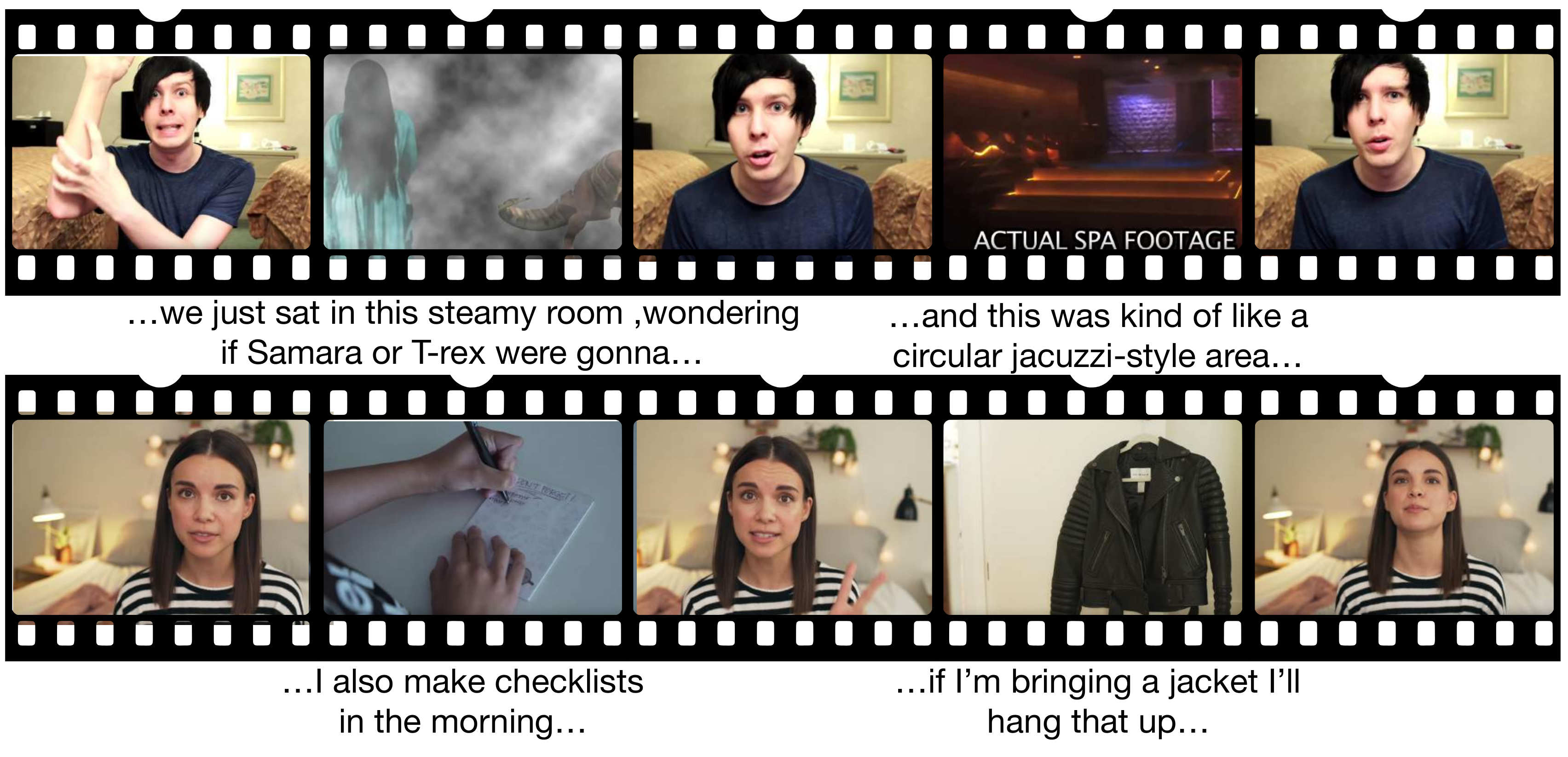}
    \caption{Screenshots from example video blogs show A-roll (talking-head of the main speaker) interspersed with B-roll (supplementary footage). In the top example, the narrator is describing his experience in a roman spa bath, and in the bottom example, the narrator is explaining her morning routine. They both leverage B-roll to support their story.}
    \label{fig:videoblog}
\end{figure}

In this work, we consider B-roll editing of vlogs, or video blogs, which are informal, journalistic documentation videos produced by so-called vloggers and are becoming an especially popular genre in social media with over 40\% of internet users watching vlogs on the web \cite{vlogs}. Vlogs usually consist of a self-made recording of the vlogger talking about a topic of interest, including personal stories, educational material or marketing. See Figure~\ref{fig:videoblog} for examples of vlogs with B-roll.

Producing engaging vlogs with B-roll has two different challenges: (1) determining which B-roll video to insert and where to insert it into the A-roll, and (2) actually performing the B-roll insertion.
In timeline-based video editors, this process often involves watching the video multiple times and experimenting with alternative choices of B-roll, which can be time consuming. Furthermore, novices often have difficulty determining how to place B-roll most effectively, which is often due to lack of experience. While professional production teams with trained experts have the advantage of resources and experience, existing video editing tools and services that assist novices (e.g., \cite{berthouzoz2012tools,pavel2016vidcrit}) provide limited support for editing videos with B-roll. 

To address the above challenges, we present B-Script, a transcript-based video editing tool that helps users insert B-Roll into the main video. 
Figure~\ref{fig:teaser} shows an overview of the B-Script system. 
In B-Script, users interact with the transcript to navigate the video, and search for diverse B-roll footage, all in one interface. 
Users insert B-roll and can modify the position and duration of the insertions using the transcript (Figure~\ref{fig:teaser}), which allows for easy alignment of B-roll with the content.
B-Script also provides automatic recommendations based on the transcript for relevant content and positions of B-roll insertion (Figure~\ref{fig:teaser}), supporting novices in finding good positions and content for B-roll.
The recommendation algorithm is trained on a new data set which we collected from 115 expert video editors.
This approach distills the experience of video editing experts into an easy-to-use interface, thereby making the authoring of videos that fit the style of successful vlogs more accessible to everyone.

In summary, we present two contributions:
\begin{itemize}
\item \textbf{B-Script, a transcript-based video editing interface.} The systems allows users to easily insert and align B-roll with video content, integrates an easy-to-use search interface for B-roll footage, and provides automatic recommendations for B-roll insertion.
\item \textbf{An automatic recommendation algorithm} that suggests B-roll content and positions for an input video. The algorithm is based on the transcript of the video.
\end{itemize}

In a user study with 110 novice video editors, we evaluated the effectiveness of our transcript-based interface and recommendation algorithm.
We compared three types of interface conditions: (1) a timeline-based interface, (2) a transcript-based interface without recommendations and (3) a transcript-based interface with recommendations. 
We also surveyed external ratings of the produced videos. 
We found that users find the transcript-based interface easier to use and more efficient compared to a timeline-based interface. 
Users also produced more engaging videos when provided with recommendations for B-roll insertions. Finally, our automatic recommendation algorithm provided recommendations of quality comparable to manually generated recommendations from experts. A demo of the system can be found at \texttt{\url{berndhuber.github.io/bscript}}.

\section{Background and Related Work}
Our work continues existing lines of research in video editing related to transcript-based editing and computational cinematography. 

\subsection{Transcript-based Editing}
Prior research has found that content-focused editing of media, such as transcript-based interfaces, can help people more effectively accomplish video and audio editing
\cite{rubin2013content,berthouzoz2012tools,casares2002simplifying,pavel2015sceneskim,shin2015visual,shin2016dynamic}. 
Casares et al. ~\cite{casares2002simplifying}, and more recently, Berthouzoz et al. ~\cite{berthouzoz2012tools} present video editing systems that include video aligned transcript editors to enable more efficient navigation and editing of video. 

Automatic location suggestion for B-roll content, as well as the B-roll browsing tool, are important novel elements in the design of B-script. QuickCut also uses a text-based interface for video editing~\cite{truong2016quickcut}. However, in it, a user provides the annotations for the video, and QuickCut uses this annotation vocabulary to generate a final narrated video composition. In our system, B-script searches over "in-the-wild" metadata, and B-script provides recommendations for content. Shin et al. ~\cite{shin2015visual} edit blackboard-style lectures into a readable representation combining visuals with readable transcripts. Vidcrit enables the review of an edited video with multimodal annotations as overlay on the transcript of the video \cite{pavel2016vidcrit}. We use these insights to build a tool that supports B-roll video editing, given its challenges of finding the right content and position for B-roll, as well as inserting B-roll.

\subsection{Computational Cinematography}
Computational assistance in such creative tasks as video and audio editing has been previously proven effective in a number of situations~\cite{grabler2009generating,iarussi2013drawing,wang2014evertutor,lee2017automatically}. Video editing assistants often rely on video metadata such as event segmentation ~\cite{lu2013story}, gaze data ~\cite{jain2015gaze}, multiple-scene shots~\cite{arev2014automatic}, camera settings ~\cite{davis2003editing}, semantic annotations and camera motions ~\cite{zsombori2011automatic,girgensohn2000semi}. Lee et al. \cite{lee2017automatically} develop a language based recommendation system for creating visually enhanced audio podcasts. Existing industry-level systems such as Wibbitz \cite{wibbitz} or Wochit \cite{wochit} help video content creators to tell visual stories, including automated content suggestions. In contrast, our tool allows users to focus on the video content in the transcript during editing via transcript-based editing and recommendations, which directly supports B-roll video editing. For developing B-Script, we build a recommendation system based entirely on the transcript of the video, using expert-annotated data for recommending video edits to users. By integrating B-roll search, and recommendations, into the workflow of B-roll insertion, our approach better supports this video editing process.

\subsection{Cinematographic Guidelines about B-Roll Insertion}
Integrating B-roll is a widely used technique in film editing for documentaries, reality shows or news coverage. 
Conventions of cinematography suggest several guidelines about B-roll editing or making cuts between shots in general. 
For example, it is recommended to avoid cuts when the speaker is in the middle of talking or gesturing actively because such cuts disrupt the audio-visual flow~\cite{o2009invisible}. 
Instead, a good practice is to place the cuts where there are natural pauses in the flow.
Sentence endings or eye blinks can be important cues for indicating these pauses ~\cite{murch2001blink}.
On the other hand, in practice, vlogs are usually created with low production value (often with a single smartphone) and do not necessarily follow the traditional principles of professional cinematography~\cite{schmittauer2017vlog}. 
For instance, in vlogs, it is common to have frequent jump cuts or rapid shot changes to make the footage faster pace. 
Instead of relying on a set of rules to determine the content and positions of B-Roll, we model our recommendation algorithm based on analysis of popular vlogs from YouTube (Section~\ref{systemdesign}) and a database of expert annotated video edits that we collect for this purpose (Section~\ref{recommendationsystem}).

\section{Formative Analyses}
\label{formative}
To learn about current practices of successful vlogs and to inform the design of B-Script, we analyzed a set of 1,100 popular vlogs on YouTube. We also designed a task for video editing experts to integrate B-roll into raw, vlog-style talking-head videos, and examined the collected data. 

\subsection{Analysis of Popular Vlogs}
To inform the design of B-Script, we analyzed a set of 1,100 popular vlogs from 4 different YouTube channels, each created by a different vlogger. Each vlog channel had a minimum of 1M followers. We selected YouTube channels based on popularity, usage of B-roll, and presence of talking-head style. We automatically extracted the B-roll from each video and analyzed its usage patterns.
We use a segmentation approach similar to~\cite{biel2011vlogsense}, which uses a content-based scene segmentation combined with face recognition to classify scenes into talking-head A-roll vs. B-roll.

\begin{figure}[t]
    \centering
    \includegraphics[width=0.59\linewidth]{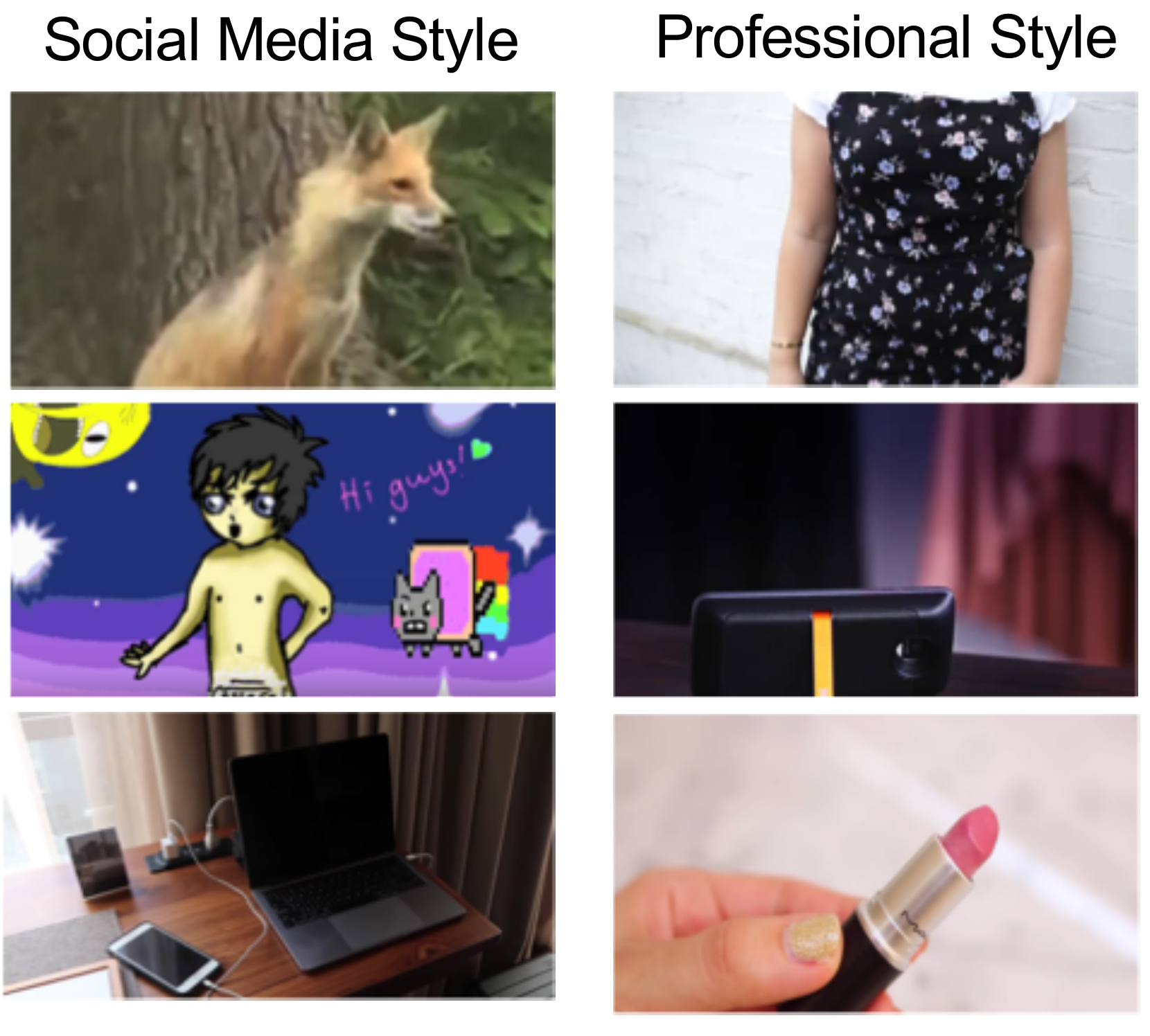}
    \caption{Examples of B-roll from a qualitative review of the B-roll in 1,100 popular vlog videos. B-roll could generally grouped into Social Media Style such as memes or casual shots, vs. Professional Style with high production quality and featuring specific objects, including product placements.}
    \label{fig:broll_categories}
    \vspace{-15pt}
\end{figure}

First, we qualitatively categorized B-roll from a randomly selected subset of 90 videos (Fig.~\ref{fig:broll_categories}) by content type and production quality in an exploratory manner. For this analysis, we extracted B-roll screenshots, aligned with corresponding transcripts.
In this process, we observed that B-roll can be grouped generally into \emph{Social Media Style} or lower production quality (e.g., screenshots and GIFs) and \emph{Professional Style} or higher production quality (e.g., professional stock video and high quality product placement).
Figure \ref{fig:broll_categories} shows examples of B-roll that we extracted in our analysis. Analysis of the aligned screenshots also revealed connections between B-roll and transcript (for example an object mentioned in the transcript may be also visible in the B-roll). This finding aligns with cinematographic guidelines on enhancing video stories with the support of B-roll~\cite{murch2001blink}.

In our quantitative analysis of the 1,100 videos, we found that B-roll clips are typically between 0.5 to 8 seconds long and evenly distributed throughout the videos. The median time between the end of one B-roll and the start of the next B-roll was 9 seconds.

\begin{figure}[t]
\centering
\includegraphics[width=.95\columnwidth]{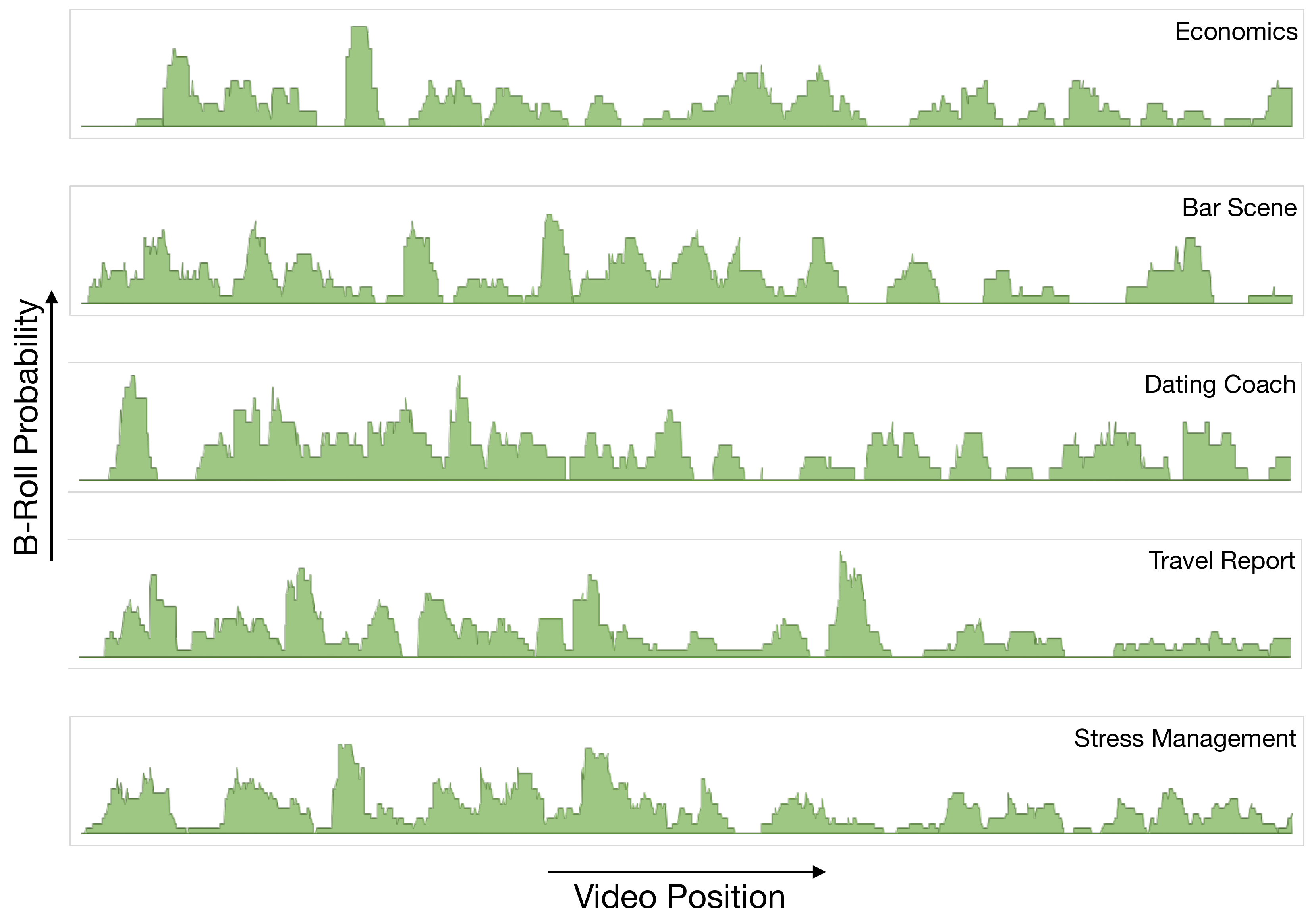}
\caption{Probabilities for insertions of B-roll for the 5 different A-roll that we used to collect expert labelled videos. Probabilities are computed by adding up, for each moment in time, the number of inserted B-roll clips, and then normalizing over the number of edited videos.
}
\label{fig:probs_broll_insertions}
\end{figure}

\subsection{Collection and Analysis of Expert Annotated Database}

To further inform the design of our recommendation system, we collected a database of expert-edited videos with B-roll. Our goal was to understand whether expert video editors inserted B-roll in a consistent manner with each other. Specifically, we wanted to observe whether there are specific locations in the video that expert editors are more likely to insert B-roll.

To collect our data set, we asked 115 expert video editors to edit a set of talking-head videos by inserting B-roll. We recruited the participants through \url{usertesting.com}, an online worker platform. Participants were filtered by their expertise in video editing skills. This expert filtering mechanism is provided by usertesting.com, which categorizes workers using an internal worker assessment system.
Experts had up to 30 minutes to edit one 3-minute talking-head video. They were instructed to make the video more engaging by inserting B-roll, and were recommended to insert approximately 8-10 B-roll clips, which aligns with the findings from our YouTube analysis. 
Participants used our transcript-based video editing interface without recommendations to edit the videos. (See Section \ref{systemdesign} for the description about our interface.)

For the input videos, we selected a set of 5 vlog videos. 
We chose talking-head style videos that did not contain any B-roll or scene cuts.  
Video topics were varied, covering economics, stress management, fitness, travelogue, and dating. 

For each of the 5 videos, we collected a set of 20-25 expert edited videos for a total of 115 edited videos. 
Experts replaced on average 13.2\% of the input video with B-roll and the median number of inserted B-roll was 8.
An overview of the aggregated B-roll insertion locations for each of the 5 different videos is shown in Figure \ref{fig:probs_broll_insertions}.\\

\noindent\textbf{B-roll Timing Clusters:}
The probability distributions for each of the videos suggest that there are clusters or areas of interest that are \textit{good} positions of B-roll. 
To quantify, how much of an overlap there actually is between different experts' annotations, we compute the average over the pairwise Jaccard coefficient. 
The Jaccard coefficient between two sample sets is defined as the size of their intersection divided by the size their union, and measures the similarity and diversity of the sample sets. For example, if two edited videos $V_a$ and $V_b$ had a set of inserted B-roll $B_a$ and $B_b$ respectively, the Jaccard coefficient would be computed as:

\begin{equation}
    J(V_a,V_b) = \frac{|B_a \land B_b|}{|B_a|+|B_b|-|B_a \land B_b|}
\end{equation}

For each input video, we take all pairs of expert-annotated videos and average their Jaccard coefficients. Then, we average the coefficients across the 5 input videos to get a final score. (Table \ref{tab:jacc}). We compare the overall coefficient with the Jaccard coefficient that would correspond to random insertions of B-roll.

\begin{table}[t]
\small
\begin{tabular}{lcc}
\toprule
& \textbf{Expert} & \textbf{Random} \\ \midrule
\textbf{Average Jaccard Coefficient (\%)} & 15.2 (2.6)            & 7.8             \\ 
\bottomrule
\vspace{-.2cm}
\end{tabular}
\caption{Mean and standard deviation of the pairwise Jaccard coefficient between the different video edits, averaged over the 5 different A-Roll. A higher Jaccard coefficient means that more \% of B-roll insertions overlap between the different video edits.}
\vspace{-0.2cm}
\label{tab:jacc}
\end{table}

The average amount of overlap is nearly twice as high between expert edits compared to between two random insertions. This suggests that there is some agreement between expert editors about \textit{good} locations to insert B-roll.\\

\noindent\textbf{B-roll Content Clusters:} To further investigate the relationship between words in the video transcript and the B-roll insertions, we looked at the query words that experts used to insert each B-roll. 
Specifically, we looked at whether the query word occurred in the transcript in a local neighborhood of the inserted B-roll, and if so how the starting position of the B-roll related to the position of the query word.

This analysis showed that for 73\% of the inserted B-roll, the query word can be found in a neighborhood of 1 second before or after the starting point of the B-roll. 
This suggests that in the majority of cases, if we could predict which word in the transcript is a good query word for B-roll, we could also recommend its position in the video.

Note that we also looked at other signals from the video. Specifically, we looked at longer pauses between speech, sentence endings and eye blinks of the speaker, signals that are recommended as good cut locations in cinematographic literature~\cite{murch2001blink,schmittauer2017vlog} . However we did not find any of these signals to be as strong a predictor for B-roll positions as the query words from the transcript. While this observation may be partially attributed to the fact that we collected the data using our transcript-based interface, we also observed this trend in our YouTube vlog analysis.\\

To summarize, our analysis of expert edited videos suggest that, in general, experts agree on good locations to insert B-roll, and furthermore, these locations are closely tied to keywords in the transcript. These facts, along with the observations from the YouTube vlog analysis informed our design of the interface (Section~\ref{systemdesign}) and the recommendation algorithm (Section~\ref{recommendationsystem}). 

\section{B-Script Interface} 
\label{systemdesign}
Our interface includes features that help editors navigate through the main video, search for relevant B-roll, insert it, and modify its duration. 
It also provides recommendations about good positions and content for B-roll. 
This section presents these features in the context of an example editing session, while the next section explains the algorithm behind the recommendation engine. 

\subsection{Video Navigation} 
Our formative analyses showed that editors choose B-roll content based on the concepts mentioned in the main video's narration. 
To facilitate this task, B-Script provides an interactive transcript as the main video navigation interface. 
The time-aligned transcript is obtained from an automatic transcription service\footnote{www.speechmatics.com/}. 
In our example, the editor uses the
transcript to navigate through the video story and identifies a text segment (\textit{``happiness''}) for which she wants to insert a B-roll (Figure~\ref{fig:teaser}A).

\subsection{B-roll Search Panel} 
Vloggers often obtain B-roll footage from online resources. As described in our analysis of popular YouTube vlogs, depending on the desired tone of the vlog, different styles of B-roll are obtained from different sources (Figure~\ref{fig:broll_categories}). To facilitate the B-roll search task, our interface integrates a search panel and provides two types of search results: \emph{Social Media Style} B-roll queried from Giphy \cite{giphy} and \emph{Professional Style} B-roll queried from Adobe Stock \cite{adobestock} (Figure~\ref{fig:teaser}. Note that in practice any external video database could be plugged in for querying.

In our example, the editor right-clicks on a word (\textit{``happiness''}) in the transcript to search for relevant footage with that query word. Alternatively, users can directly type in the query word (Figure~\ref{fig:teaser}A). The editor can then scroll through the list of search results from each category.

\subsection{B-roll Insertion and Manipulation}
When the editor finds a desired B-roll clip from the search result, she can insert it by clicking on the ``Add Now'' button below the B-roll. The B-roll is inserted into the current position of the video with a default duration which corresponds to the duration of the B-roll footage. We set the minimum duration to 0.5 seconds and the maximum duration to 8 seconds, in accordance with the typical range of B-roll duration from our YouTube vlog analysis (see Section~\ref{formative}). 

The duration of the B-roll is indicated with a green highlight over the transcript, and the content of the B-roll is indicated with a thumbnail of the B-roll at the beginning of the text. The editor can review the insertion by playing the video in the main video panel. 
To change the starting position of the inserted B-roll, the editor can drag the thumbnail and drop it elsewhere in the transcript. The duration of the B-roll can be modified by dragging the end of the B-roll insertion overlay accordingly. 
Making the video shorter than the actual length of the B-roll will cut the end, while making the video longer will loop the B-roll video. Finally, B-roll insertions can be deleted by dragging the B-roll thumbnail onto a trashcan icon at the right top of the screen (not shown in Figure~\ref{fig:teaser}). In our example, the editor inserts a B-roll related to the query word \textit{happiness} and modifies its position and duration to overlay the part of the narration that says \textit{``And that's what works for me.''}. (Figure~\ref{fig:teaser}B)

\subsection{Recommendation User Interface}
In order to assist editors with their creative decisions, our interface can also provide automatic recommendations for B-roll position and content. Recommendations are shown as yellow highlighted overlays over the video transcript. 
A button at the right side of each overlay indicates the query word associated with the recommendation (Figure~\ref{fig:teaser}C). 
The editor can click on the highlighted transcript text or the query word to retrieve B-roll videos with the search query. If the editor likes one of the search results, she can click the ``Add Now'' button to insert the video over the highlighted portion of the transcript. We chose this mixed-initiative approach so that editors have the option to easily accept or reject the recommendation and to select their own style of B-roll from the query results, while the recommendations serve to facilitate the process. 
The following section describes the design and technical evaluation of the recommendation algorithm.

\section{B-Roll Recommendation Algorithm} 
\label{recommendationsystem}
Our analysis of expert-edited videos in Section~\ref{formative} suggests that there are positions in videos that experts choose to insert B-roll significantly more often, and that these locations are closely tied to keywords in the video transcript. This section describes the development and evaluation of our B-roll recommendation algorithm based on these findings. The goal of the algorithm is to predict the start word for B-roll insertion. We return the start word as the recommended search term for B-roll insertion as it often corresponds to a good candidate keyword for B-roll search.

\begin{figure}[t]
\centering
\includegraphics[width=0.75\columnwidth]{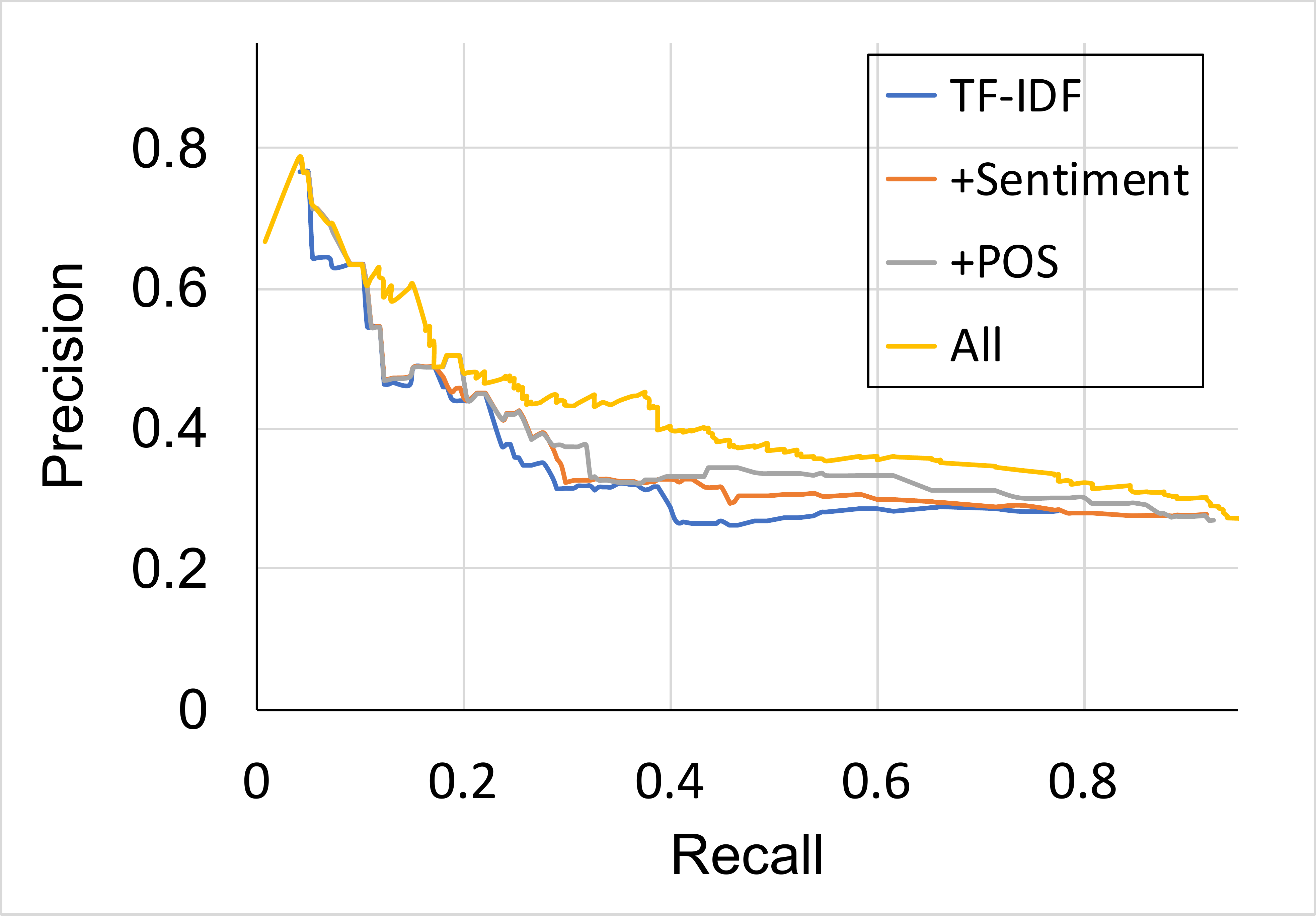}
\caption{Precision-Recall curve for our algorithm. Starting with TF-IDF feature only, the classifier achieves better performance by adding more features.}
\label{fig:roc}
\vspace{-10pt}
\end{figure}

We analyze the words in the transcript by extracting (1) TF-IDF scores, (2) word sentiment, (3) its part of speech (POS), and (4) the number of occurrences of the word within the transcript. For all of our transcript analyses, stop-words (i.e., words that occur very often in English language such as `the' or 'and') are excluded.

To compute the TF-IDF feature vector, we used the transcripts from the full set of 1,100 videos from our YouTube analysis. For sentiment analysis, we used the sentiment classifier by Hutto et al~\cite{hutto2014vader}, which computes a continuous sentiment score from -1.0 (very negative) to 1.0 (very positive). The sentiment term models the potential impact of text sentiment on the video. POS tags were computed using the Spacy POS tagger~\cite{kiperwasser2016simple}, and were translated into numeric scores using a one-hot feature vector (i.e., 1 for the position that corresponds to the POS tag of the word and 0 elsewhere). This term models potential signals of certain parts of speech (e.g., nouns) being more likely to be a query word than others. Finally, we took into account the repeated occurrences of a given word by giving a score to each word equal to the number of previous occurrences of the same word. This term models the potential decrease in the importance of a word as it occurs repeatedly. Finally, we concatenate the four terms into a 5,033-dimensional feature vector for each word. 

Using this representation, we trained a linear classifier (SVM with a linear kernel, $C=0.01$)~\cite{suykens1999least} that predicts whether a word is a keyword or not. For training, we used the expert-annotated database for 5 videos from 115 experts. We used cross-validation, leaving out data for one video and predicting query words for that video based on the training data from the other four videos.

To evaluate the performance of our classifier, we compare the prediction results with the agreement level between human experts. Our results show that our algorithm approaches close to experts' agreement level. With a precision threshold of 0.6, our algorithm scored a total F1 score of 0.23 as compared to 0.26 for experts, with the F1 score being the harmonic mean of precision and recall. 
Detailed results of this cross-validation analysis can be found in Table \ref{tab:accuracy_query_predictions}, and the impact of the different features on the performance in Figure \ref{fig:roc}. Combining all the features achieved best performance. These results suggest that query word recommendations for B-roll can be generated algorithmically with reasonable quality based on video transcripts.

\begin{table}[]
\small
\begin{tabular}{rccc}
\toprule
                   & \textbf{Algorithm} & \textbf{Human} & \textbf{Random} \\ \midrule
\textbf{Precision} & 0.61                & 0.30      &  0.13    \\ \midrule
\textbf{Recall}    & 0.14                & 0.23      &  0.15    \\ \midrule
\textbf{F1}        & 0.23                & 0.26      &  0.14    \\ 
\bottomrule
\end{tabular}
\caption{Cross validation results for the B-roll keyword detection algorithm, compared to human-expert agreements.}
\label{tab:accuracy_query_predictions}
\vspace{-19pt}
\end{table}

\section{User Study}
Our goal is to design a system to facilitate novice vloggers to produce better videos by inserting B-roll. 
To evaluate the effectiveness of our design choices, we conducted a user study to address the following research questions:

\begin{description}
\item[RQ1] Is the transcript-based interface more effective than the baseline timeline-based interface?
\item[RQ2] Does the recommendation system help the editing task? How is this affected by recommendation quality?
\item[RQ3] Does the recommendation system help create a better outcome? How is this affected by recommendation quality?
\end{description}

\subsection{User Interface Conditions}
\begin{figure}[t]
    \centering
    \includegraphics[width=0.98\linewidth]{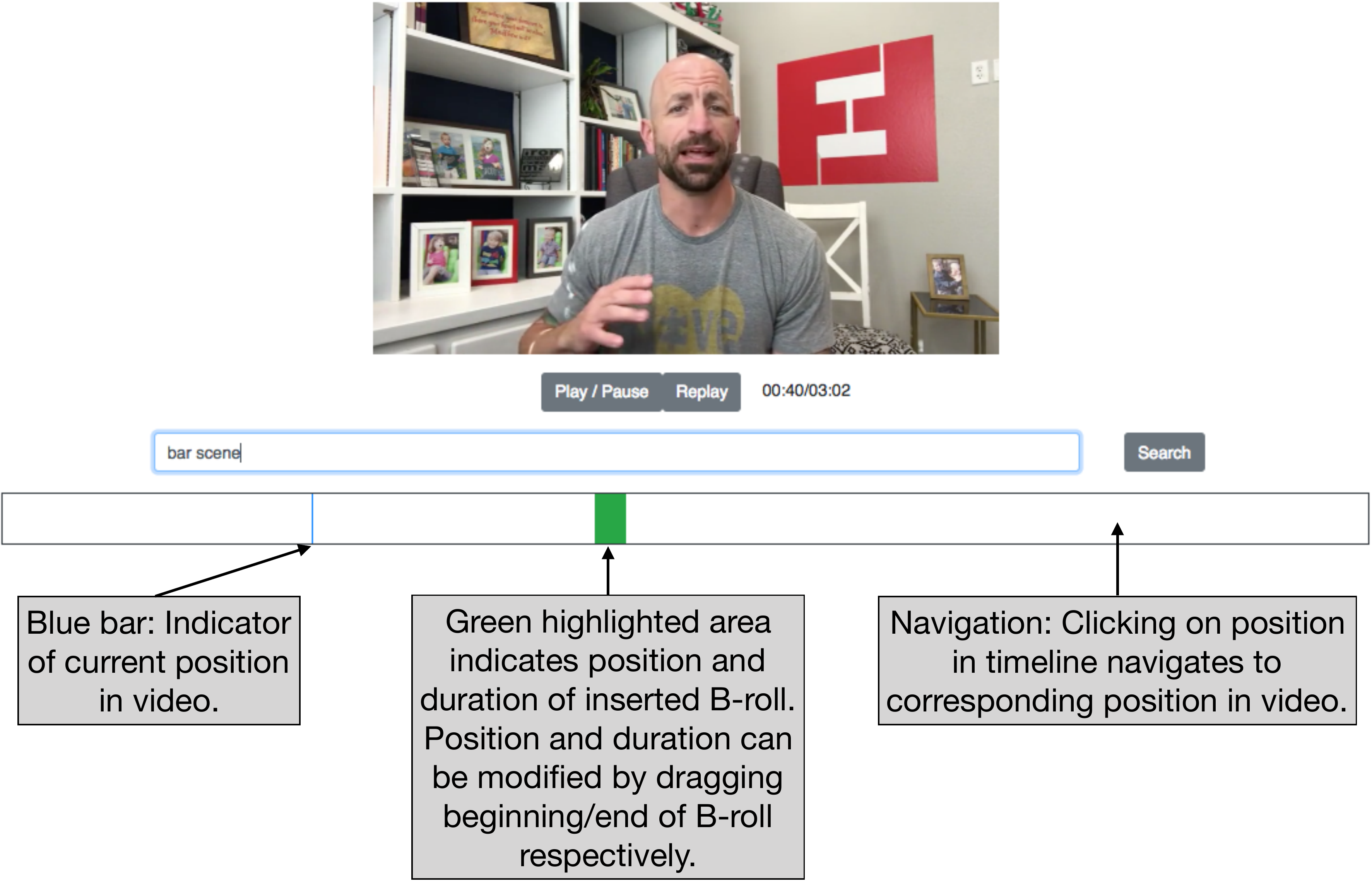}
    \caption{The baseline timeline based user interface for B-roll editing.}
    \label{fig:timeline}
    \vspace{-10pt}
\end{figure}

Our study evaluates three different user interface conditions as our main factors: a reference timeline-based system (Figure ~\ref{fig:timeline}), a transcript-based system without recommendations, and a transcript-based system with recommendations. 
The timeline-based system provides the same navigation, search and insertion capabilities as the transcript-based system, except that a timeline is used for B-roll video editing, similar to existing popular video editing tools such as iMovie \footnote{https://www.apple.com/imovie/}. We leave out the combination of timeline-based with recommendations, since we considered this out of scope for our study.

\subsection{Recommendation Conditions}
Since our goal is to evaluate the effect of the quality of recommendation on the effectiveness of our interface, as well as the quality of our recommendations algorithm, our study additionally evaluates recommendations generated by our algorithm, by experts, and interval-based recommendations. 

First, our algorithmic recommendations were generated using the leave-one-video-out approach, similar to our cross-validation evaluation, to predict query words in the transcript. 
In this case, recommended duration was set to two seconds. 

Second, we aggregated the expert annotations per video and selected areas that were above mean probability of a B-roll insertion as recommendation. 
We then selected the most frequently occurring query associated with this area as the query recommendation. 
The recommended duration of the B-roll was set to the position where the probability dropped again below the mean. 
Finally, the interval-based recommendations were generated with a basic heuristic aligned with the observations we gathered from our YouTube vlog analysis. 
The interval-based recommendations were generated so that every nine seconds, a two second B-roll recommendation occurred. 
The recommended query was taken as the first transcript word at the recommendation position. 
Table \ref{tab:recommendation_queries} shows example queries of the three different conditions.

\begin{table}[t]
\small
\begin{tabular}{p{0.23\linewidth}p{0.7\linewidth}}
\toprule
                     & \textbf{Example queries from one video}                                                                                                         \\ \midrule
\textbf{Algorithm} & mornings, care, people, people, life, passion, thing, somebody, break, guy, rest, aspects, periods, school, break, lessons, creativity, YouTube \\ \midrule
\textbf{Expert}      & stress, public figure, working, passion, happiness, lifting weights, injured, stronger, school, break, burned out, stressed, YouTube            \\ \midrule
\textbf{Random}      & I'm, unhealthy, people, always, and, that, weights, lifts, hard, I came, been going, out, and, analytics                                        \\ 
\bottomrule
\end{tabular}
\caption{Example queries for the same video with the three recommendation generation approaches.}
\label{tab:recommendation_queries}
\vspace{-21pt}
\end{table}

\subsection{Questionnaires}
All questionnaires were asked post-stimulus, after users finished one video, and before editing the next video.\\

\noindent\textbf{Task difficulty: } 
Task difficulty was evaluated with the NASA-TLX questionnaire \cite{hart1988development}. The NASA-TLX assessment is commonly used for measuring the overall difficulty of a task.
This questionnaire consists of a set of questions related to how mentally and physically demanding and stressful the task is.
Responses are recorded on likert scales between 1 and 10. 
We analyze the raw NASA-TLX score, which is computed by calculating the mean score of the six NASA-TLX questions.\\

\noindent\textbf{Helpfulness: } 
We asked users to rate the helpfulness of the search interface, the video insertion function, and the B-roll manipulation interaction (for modifying its position and duration), each at a 5-point likert scale. Specifically, every participant scored each interface feature on a scale of how much it helped her during the task (1-not helpful at all, 5-very helpful). As an overall measure, we also asked people whether they liked the interface overall.\\

\noindent\textbf{Outcome satisfaction: }
As a measure for outcome quality, we asked participants to subjectively rate how satisfied they were with the video they produced, at a 5-point likert scale.

\subsection{Participants}
We recruited participants through the usertesting.com online worker platform. Participants were pre-screened by the platform to have some level of experience with video editing, but no significant experience. This filter was applied in order to target novices, while also taking into account potential learning effects that may interfere with our study. 
Workers on the usertesting.com platform expect maximum task durations of 30 minutes.
To balance the participant workload with our experiment design, we chose to assign three, 3 minute videos per worker.
We tested this setup in a pilot study, and found that 30 minutes for three conditions was a reasonable trade-off between length of the task and number of conditions. 
We recruited 180 participants, out of which 110 finished all three video edits within the 30 minutes. In all of our data analyses, we analyze the data from those 110 participants who completed all three video edits. The vast majority of workers who dropped out did not manage to reach the third condition. While this time constraint might affect the time people spend on each task and excludes people who ran out of time, we assume this minimally affects the validity of our analyses. 

\begin{figure}[t]
    \centering
    \includegraphics[width=0.75\columnwidth]{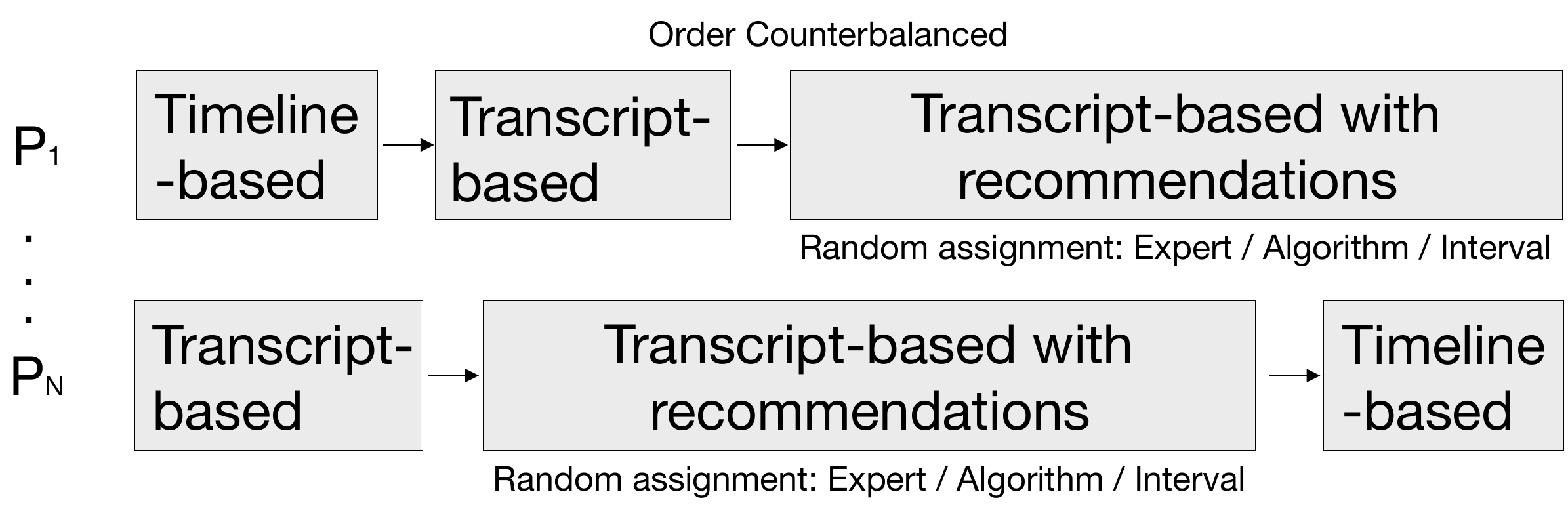}
    \caption{Experiment overview with the different conditions to evaluate transcript-based video editing, as well as the recommendation system.}
    \label{fig:experiment_design}
    \vspace{-13pt}
\end{figure}

\subsection{Procedure}
For our study, we use three of the five A-roll videos that we also used for the expert data collection task in Section~\ref{formative}. All videos and user interface conditions were complete counterbalanced in order. Workers received \$10 for the 30 minute task. Workers were instructed to make the video more engaging by inserting B-roll. They used three different user interface conditions to edit the three videos. Workers accessed B-script via a web interface. Before the start of the task, workers were asked to report basic demographic information, how good they are at video editing, and how familiar they are specifically with the B-roll video editing task.

To keep the number of conditions and the task-length appropriate for the participants, we tested the recommendation quality in a between-subject design. 
Users were randomly assigned to either expert-annotation based recommendations, our algorithmic recommendations, or interval-based recommendations. Figure \ref{fig:experiment_design} provides a detailed overview of the experiment design.

\subsection{External rating task}
In addition to asking participants about their subjective satisfaction of their own video output, we also asked external raters to rate the videos created by participants in the experiment.
To limit the number of videos to be rated, we took a randomly selected subset of 72 videos, half created using the transcript-based interface without recommendations and the remaining half created using the transcript-based interface with recommendations. 
The external raters reported basic demographic information and how frequently they used social media. 
The raters were then asked to watch a randomly assigned video for a duration of three minutes. 
After viewing the video, raters were asked \textit{how likely they would give the video an upvote} if they would see it on a video sharing platform such as YouTube, Vimeo or Facebook. The response options were on a 5-point likert scales from very unlikely (1) to very likely (5). 
We used three criteria to filter the responses: first we had a screening question about the general content of the video and filtered out responses with wrong answers; second, since we wanted to measure how appealing the video was for social media users, we filtered out data from raters who reported they never or very rarely used social media platforms; finally, we filtered out raters who reported that they do not up-vote posts on social media in general. In all, we analyze ratings from 580 raters (age M=26.5, SD=11.5, Male:56\%, Female:44\%) for the 72 videos. 

\subsection{Design and Analysis}
There were two measures of interest to investigate the effect of timeline-based vs. transcript-based video editing interface: (1) the perceived task difficulty as measured by NASA-TLX raw scores; and (2) the average time taken per B-roll insertion as a measure of relative time taken to create an outcome. For our within-subject evaluation, we use a repeated measures ANOVA, with user interface condition and trial number as factors. 

To investigate the effect of recommendations, we measure perceived helpfulness of the tool, outcome satisfaction as reported by participants, and number of inserted B-roll. For our between-subject evaluation, we compare recommendation quality and user interface type as factors. For this analysis, we leave out the timeline-based condition.

\section{Results}

\subsection{User-Edited Video Statistics}
To provide a context for the data that we collected, we first present some general statistics from the created videos.

In total, our user study consisted of 110 participants.
For the 330 videos created by those participants, the average number of B-roll clips that were inserted was M=6.8 (SD=3.7). 
Video editing sessions took on average 8.8 minutes for each video. 
The average time that participants took to insert a single B-roll was M=97.0 seconds (SD=47.1). 
The average duration of an inserted B-roll was 3.7 seconds and the average distance between two B-roll in a video was 14 seconds.

\begin{figure}[t]
\centering
\begin{subfigure}{.48\columnwidth}
\includegraphics[width=\textwidth]{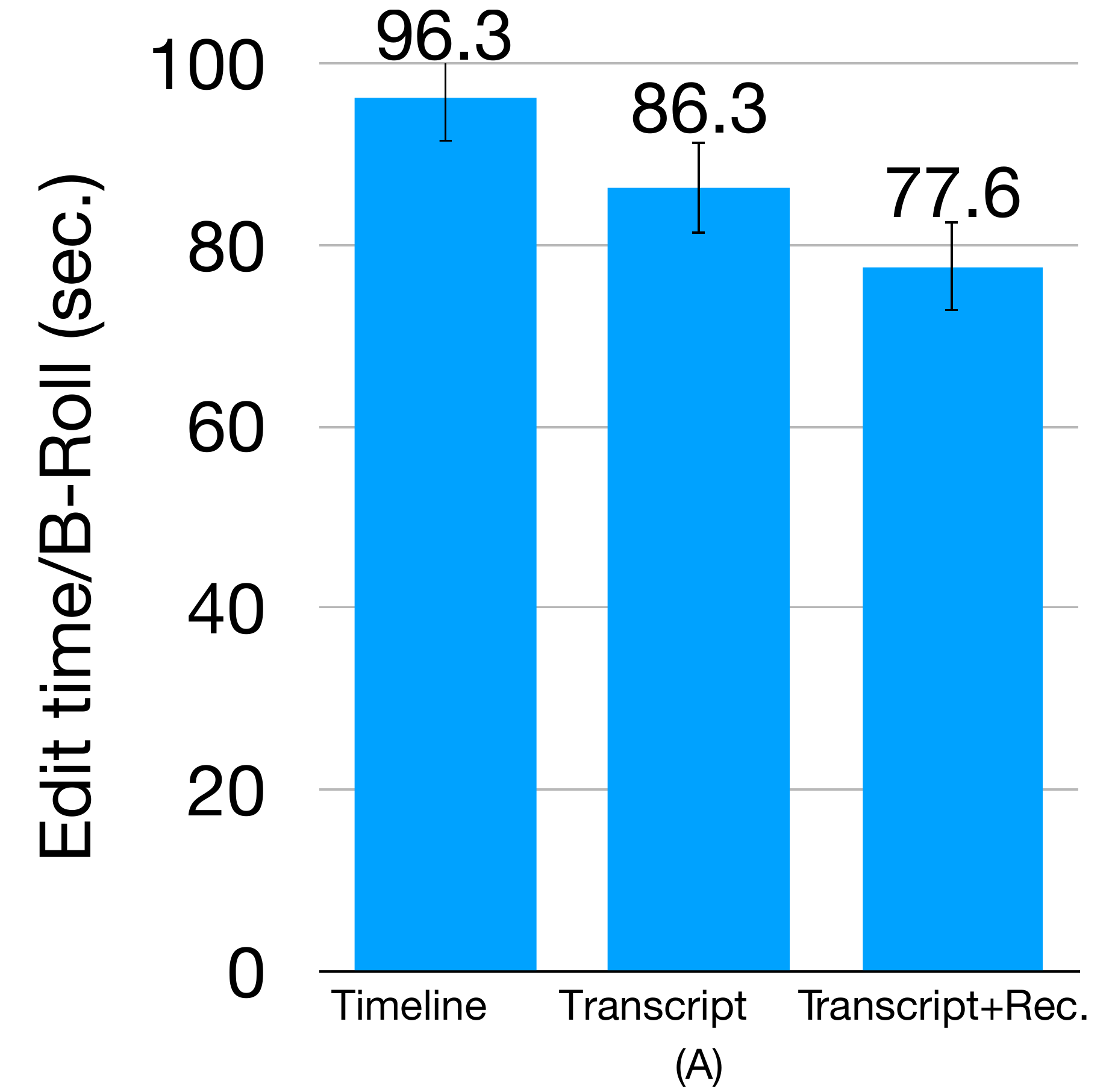}
\end{subfigure}
\begin{subfigure}{.48\columnwidth}

\includegraphics[width=\textwidth]{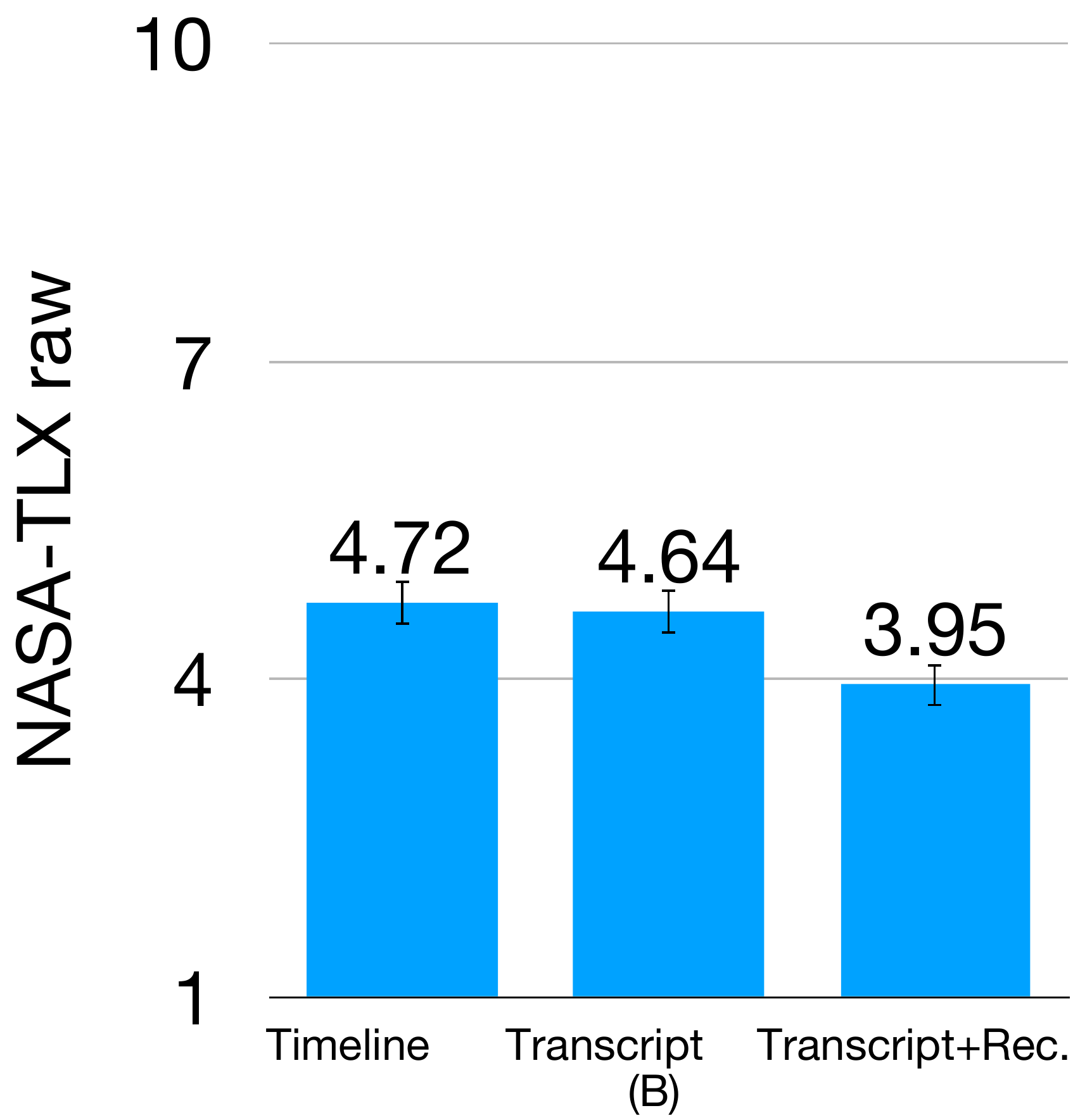}
\end{subfigure}
\caption{(A) Average duration of editing time, normalized over the number of B-roll insertions, for the three different UI conditions. (B) Raw NASA TLX scores (lower is better), for the three different UI conditions.}
\label{fig:text_vs_timeline}
\end{figure}
 
We also looked at the types of B-roll that participants selected for insertion. 
57\% of inserted B-roll came from the social media style category, and 43\% came from the professional style category. 
The median rank of the selected B-roll within the search results was 3. 
The median ranks by video category were 3 for the social media style videos and 4 for the professional style videos. 
Finally, the median number of queries that participants entered while editing the videos was 18.

\subsection{Transcript-based interface is more efficient than timeline-based interface (RQ1)}
There were two dependent variables to study the effectiveness of the B-roll video editor: (1) the average time taken per B-roll insertion as a measure of relative time taken to create an outcome; and (2) the perceived task difficulty as measured by NASA-TLX raw scores (lower means less difficult). 
Overall, the average time per B-roll insertion was 85 seconds and the average NASA-TLX score was 4.3. Figure~\ref{fig:text_vs_timeline} shows the average editing times per B-roll and raw NASA-TLX scores for the different user interfaces. In both measures, we observe an overall positive trend (less editing time and lower NASA-TLX score) for transcript-based editing. 
A repeated measures ANOVA analysis did show a statistically significant main effect of user interface type with regards to editing time per B-roll ($F(2, 328) = 2.91, p = .0217$), and NASA-TLX raw scores ($F(2, 328) = 3.87, p = .0230$). 
No significant interaction effect between trial number and user interface type was found. 
A Tukey HSD post-hoc test found the average time taken per B-roll insertion to be significantly lower for transcript-based interfaces ($t(220) = 3.07, p = 0.030$) compared to timeline-based interfaces.
No significant pairwise difference was found for NASA-TLX raw scores between the timeline-based and transcript-based interface. 

\begin{table}[t]
\small
\resizebox{.95\linewidth}{!}{
\begin{tabular}{p{0.24\columnwidth}ccc}
\toprule
                               & \textbf{Timeline} & \textbf{Transcript} & \textbf{Transcript + Rec} \\ \midrule
\textbf{Search\newline Feature}        & 3.8 (1.4)               & 4.2  (1.2)         & 4.4  (1.0)                        \\ \midrule
\textbf{Position\newline Manipulation} & 4.0   (1.3)            & 4.1   (1.3)        & 4.1  (1.3)                        \\ \midrule
\textbf{Duration\newline Manipulation} & 4.2   (1.2)            & 3.9  (1.4)         & 3.9  (1.4)                        \\ 
\bottomrule
\end{tabular}
}
\caption{Mean and standard deviation of B-Script features rated by participants based on helpfulness for the task, on a scale from 1 to 5, with 5 being most helpful. In the transcript-based editing interface, users rated the search interface higher than the video manipulation features. In the timeline-based interface, the order was reversed. }
\label{fig:ux_features}
\vspace{-27pt}
\end{table}

\subsection{Transcript-based interface is as easy to use as timeline-based interface (RQ1)}
Users rated the helpfulness of different features of the interfaces (Table \ref{fig:ux_features}).
The average user rating for the transcript-based interface was as least as high as the average rating for the timeline-based interface. 
In the transcript-based editing interface, users rated the search interface higher than the video manipulation features. 
In the timeline-based interface, the order was reversed. 
Overall, the search feature showed the largest improvement in the transcript-based interfaces.

\subsection{High quality recommendations are helpful (RQ2)} 

\begin{figure*}[t]
    \centering
    \begin{subfigure}{.32\textwidth}
    \includegraphics[width=0.95\columnwidth]{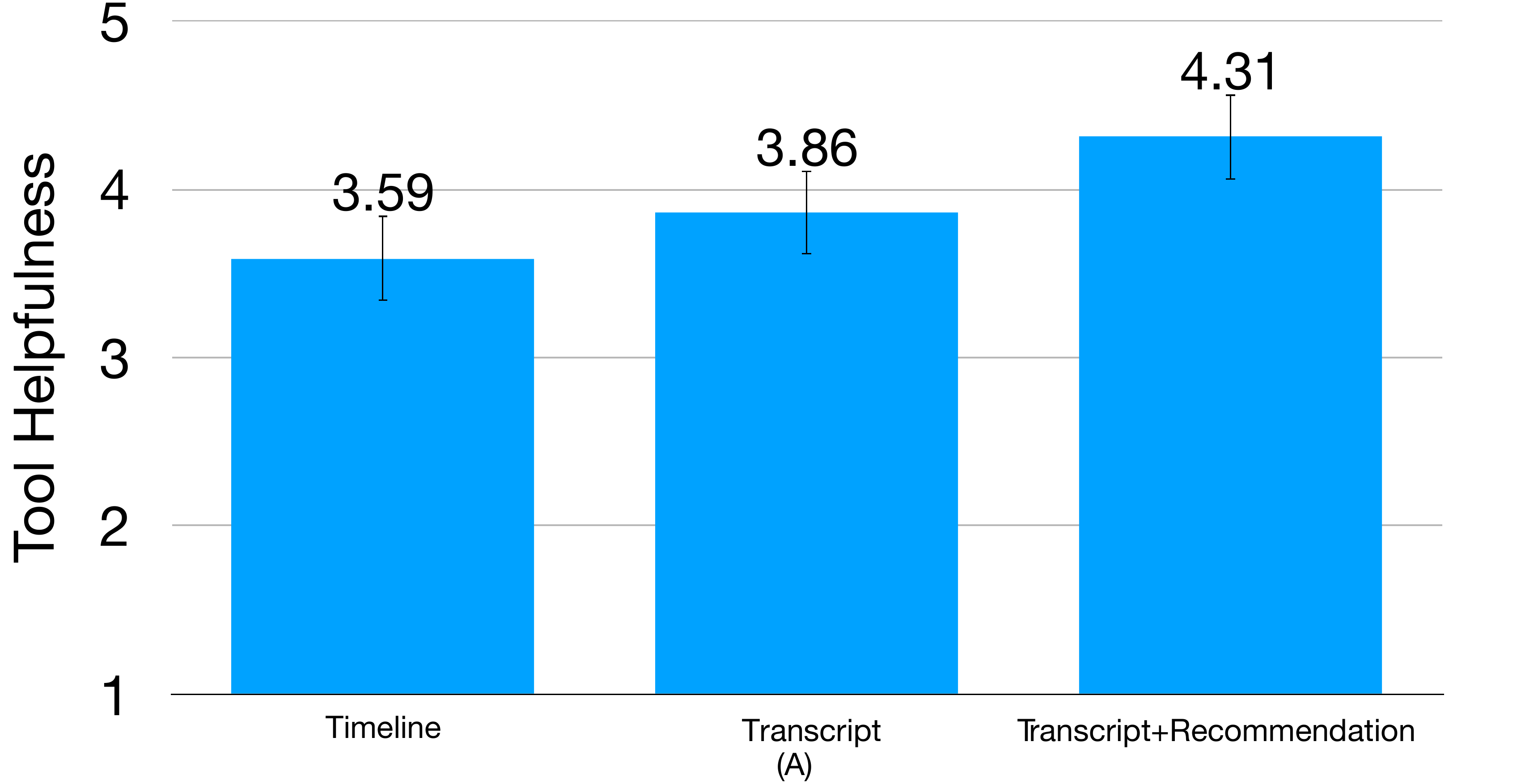}
    \end{subfigure}
    \begin{subfigure}{.32\textwidth}
    \includegraphics[width=0.95\columnwidth]{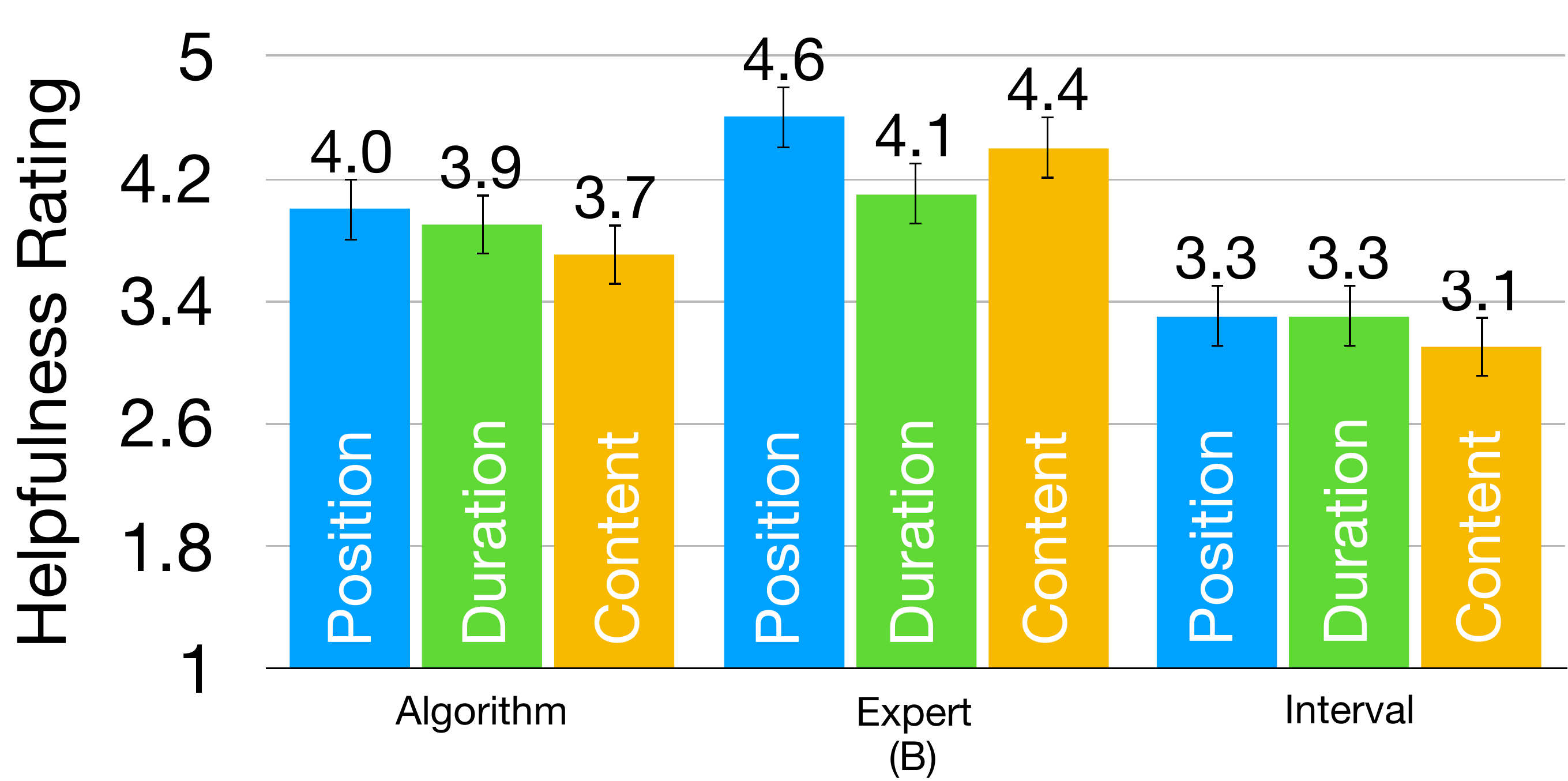}
    \end{subfigure}
    \begin{subfigure}{.32\textwidth}
    \includegraphics[width=0.95\columnwidth]{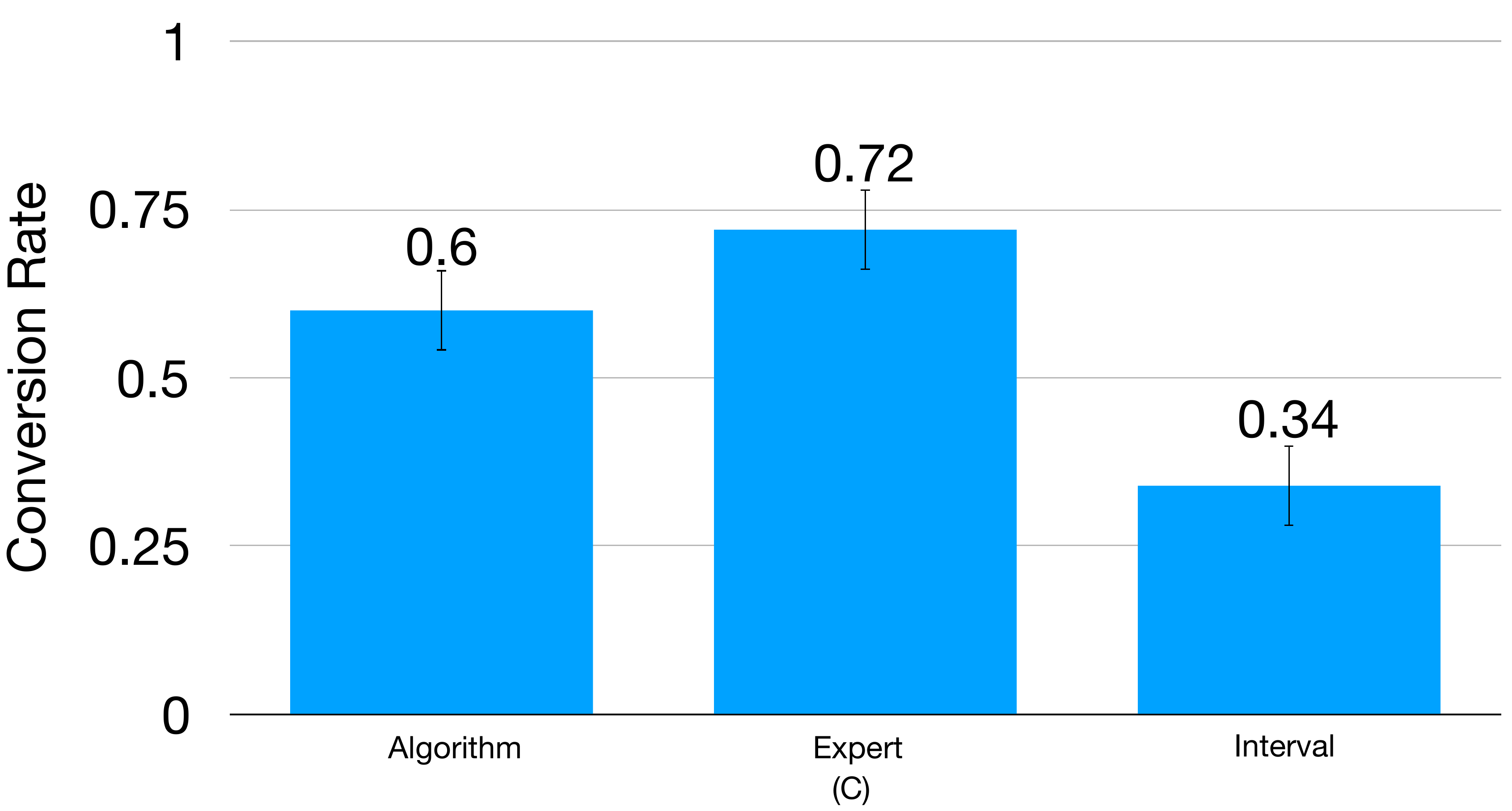}
    \end{subfigure}
    \caption{Recommendation are generally helpful for the B-roll video editing task. (A) General helpfulness of the tool to insert B-roll, as reported by the participants. (B) Helpfulness of the different recommendation features (position, duration and suggested query), as reported by the participants. (C) Recommendation conversion rates, which was computed as the fraction of inserted B-roll that was taken from the recommendations. }
    \label{fig:text_vs_recommendation}
\end{figure*}

To answer how helpful recommendations are, we looked at the perceived helpfulness of the tool as reported by the users. 
The average level of perceived helpfulness was 3.9 on a likert scale from 1-5. 
An ANOVA analysis shows a statistically significant effect with regards to interface type ($F(2, 176) = 14.4, p = .01$). 
The mean helpfulness ratings are shown in Figure ~\ref{fig:text_vs_recommendation}. Users rated the transcript-based interface with recommendation as the most useful. 

There were two dependent variables to study the effect of recommendation quality on the user experience: (1) the perceived helpfulness of the specific features of the recommendations including duration, position and content; and (2) the conversion rate of the recommendations as measured the fraction of recommendations that were directly converted into B-roll insertions. 
An ANOVA analysis showed a statistically significant main effect for both recommendation helpfulness ratings ($F(2, 108) = 24.04, p = <.0001$) and conversion rates ($F(2, 108) = 11.72, p = <.0001$). 
The mean ratings can be seen in Figure ~\ref{fig:text_vs_recommendation}. Users rated the expert recommendations as most useful, followed by our algorithmic recommendations. The interval-based recommendations were rated as least helpful.

\begin{figure}[ht]
    \centering
    \begin{subfigure}{.48\columnwidth}
    \includegraphics[width=0.95\columnwidth]{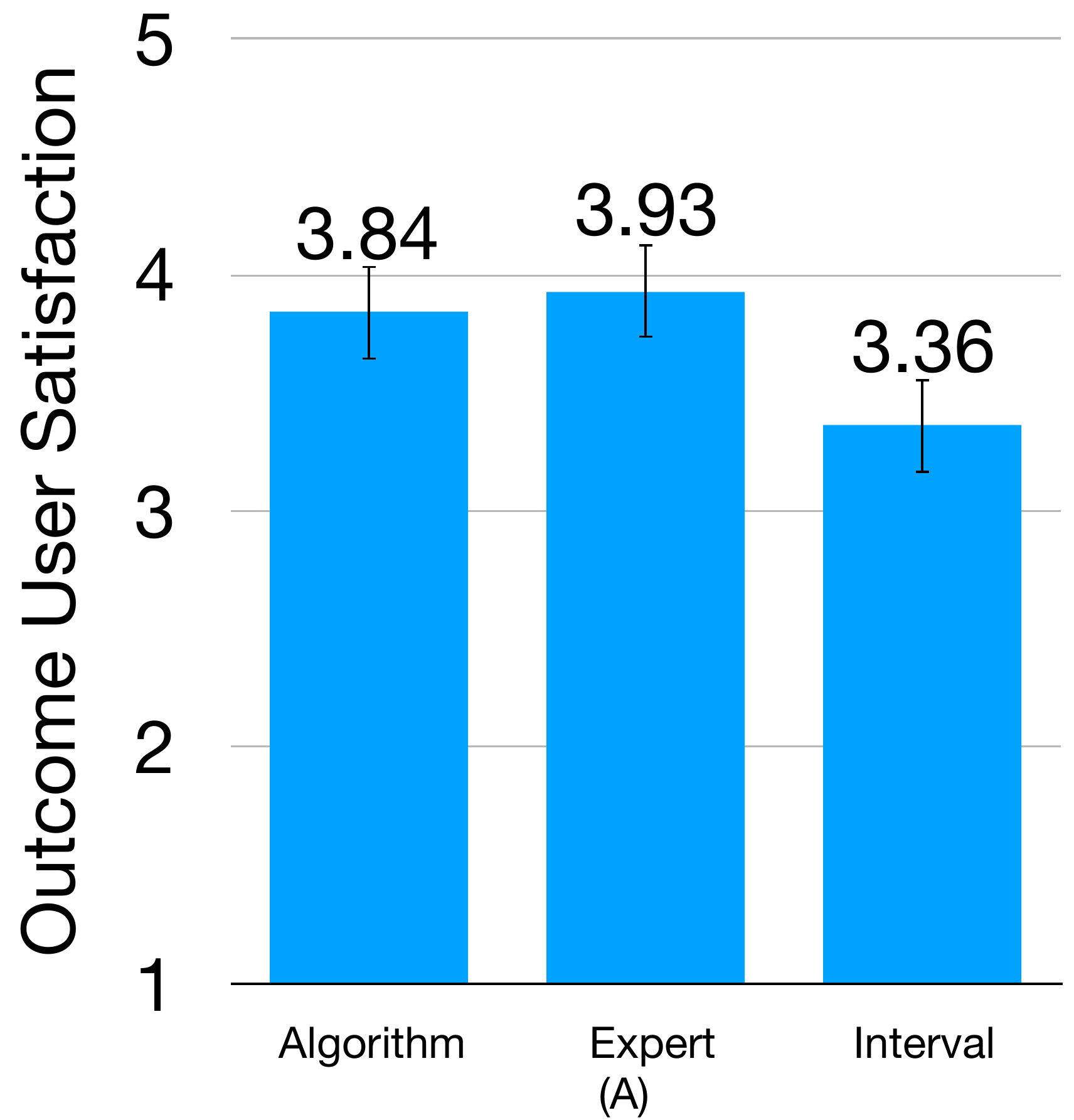}
    \end{subfigure}
    \begin{subfigure}{.48\columnwidth}
    \includegraphics[width=0.95\columnwidth]{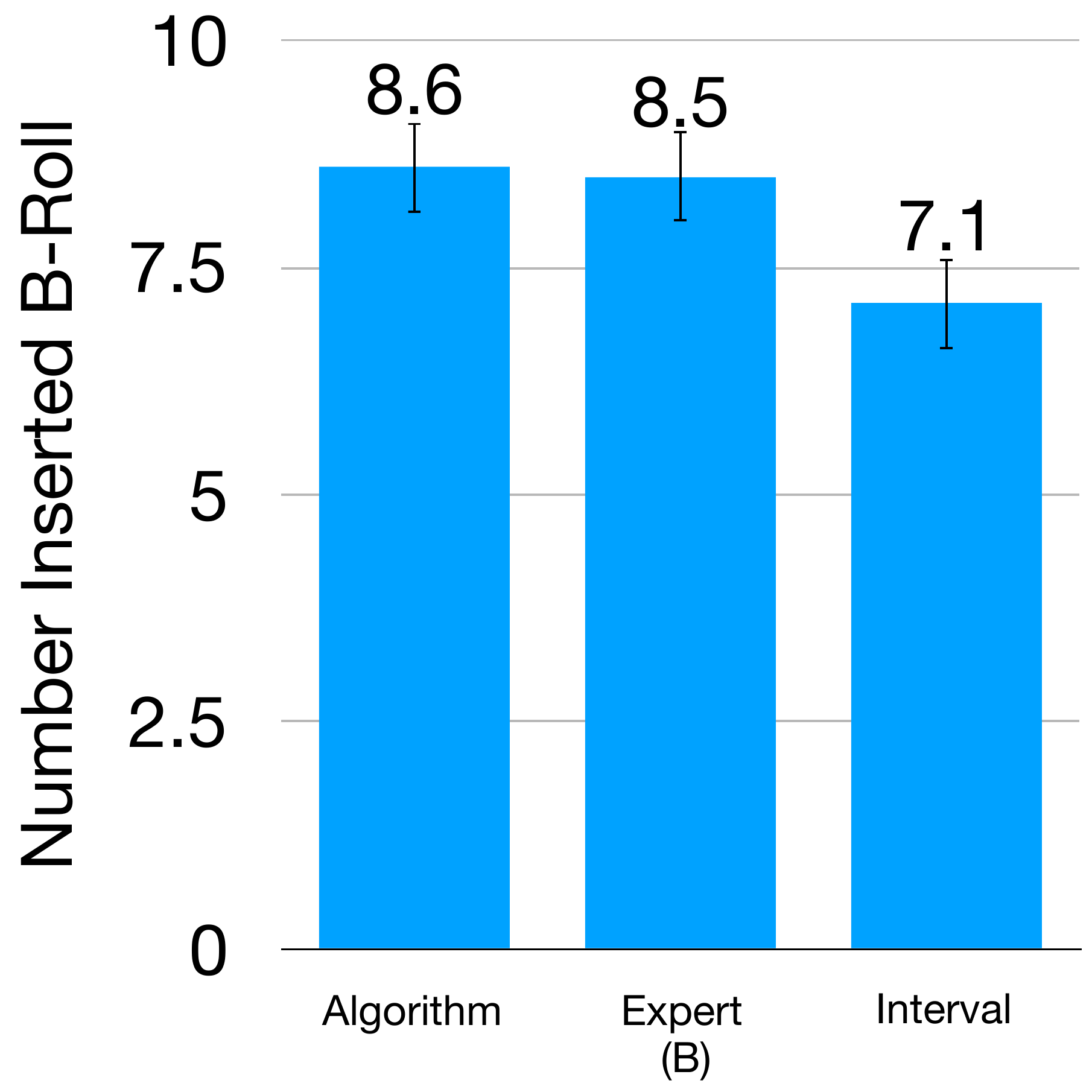}
    \end{subfigure}
    \caption{Our results suggest that videos created with recommendations are better. (A) Average user outcome satisfaction of the three different interfaces. (B) Average number of inserted B-roll of the three different interfaces. }
    \label{fig:outcome_analyses}
    \vspace{-10pt}
\end{figure}

\subsection{High quality recommendations produce better videos (RQ3)}

Users in the recommendation condition reported to be significantly more satisfied with their outcome ($F(2, 176) = 10.5, p = .02$). Furthermore, users inserted significantly more B-roll into the videos in the recommendation conditions ($F(2, 176) = 16.1, p = .001$).

An ANOVA analysis showed a statistically significant main effect of recommendation quality with regards to the number of B-roll inserted ($F(2, 108) = 1.8, p = .023$) and satisfaction of users with the outcome ($F(2, 108) = 3.20, p = .043$). 
The mean ratings can be seen in Figure~\ref{fig:outcome_analyses}. In general, users inserted more B-roll and were more satisfied with the outcomes created using the expert recommendations and our algorithmic recommendations compared to the interval-based recommendations. 
External raters were also significantly more likely to upvote the videos that were created with the user interface with recommendations (M=2.9) as compared to the videos that were created without recommendations (M=2.6). An ANOVA analysis showed that the difference is statistically significant ($F(1, 579) = 4.92, p = .027$).

\section{Discussion and Conclusion}

Our results suggest that transcript-based video editing with recommendations is a promising way to effectively edit videos with B-roll. The transcript-based interface significantly reduced the editing time per inserted B-roll, and also reduced the perceived task difficulty compared to the timeline-based interface.
Furthermore, users rated the system as significantly more helpful when recommendations for B-roll insertions were shown. 
Our system also showed significant improvements on the quality of the produced videos, both subjectively (rated by creators and external raters) and quantitatively (measured by number of B-roll insertions).
In the condition with the recommendations, users were significantly more satisfied with the outcome, as reported by themselves, and users inserted significantly more B-roll than when recommendations were not shown.
Finally, external raters rated the videos generated with our recommendation system significantly better and more likely to give upvotes to, if the video was on social media.

We also observed that user ratings significantly varied with recommendation quality. 
This suggests that the quality of recommendations is important, and that users do not just click on the recommendations because they are there. 
One limitation of our work is that the expert dataset that we collected was still relatively small. 
We expect that our recommendation quality would improve by gathering more labelled data from users, and that a larger data set will lead to a better understanding of more generalizable signals with further insights. 
Nonetheless, the improved outcome quality with our algorithmic recommendations suggest that even our current system can be useful in practice.

As future work, we imagine that our approach can be extended to other types of videos and editing tasks.
In this work we focused on a relatively narrow genre of videos in the video blog domain. 
To understand how transcript-based interfaces affect video production outcomes in general, studies in such different domains as education or news are required. 
For example, in such educational settings as MOOCS, where video creation is a crucial part, transcript-based video editing may be a tool that helps instructors create more engaging videos. 
Furthermore, while we did not find significant relationships between B-Roll insertions and visual or audio cues, an interesting future direction might be to explore how computational cinematography benefits from showing such signals as an overlay on the text, similar to Rubin et al.'s work on audio story editing~\cite{rubin2013content}.
We also focused on style of B-roll rather than the source of the content to enable styles similar to successful vlogs. Leveraging personal content as B-roll is an interesting possible extension of our tool, and the categories presented in the B-roll browser can be easily extended.
Finally, we note that our system currently supports only simple overlay video insertions. 
One could imagine scenarios where, for example, the B-roll appears as an inset on top of the A-roll or where the A-roll audio also needs to be replaced.

\bibliographystyle{ACM-Reference-Format}
\balance
\bibliography{paper}

\end{document}